\documentclass[conference]{IEEEtran}
\IEEEoverridecommandlockouts

\usepackage{cite}
\usepackage{hyperref}
\usepackage{makecell} 
\usepackage{amsmath,amssymb,amsfonts}
\usepackage{algorithmic}
\usepackage{graphicx}
\usepackage{balance}
\usepackage{textcomp}
\usepackage{colortbl}
\usepackage[table]{xcolor} 
\usepackage[most]{tcolorbox}
\usepackage{xcolor}
\usepackage{enumitem}
\def\BibTeX{{\rm B\kern-.05em{\sc i\kern-.025em b}\kern-.08em
    T\kern-.1667em\lower.7ex\hbox{E}\kern-.125emX}}

\tcbset{textmarker/.style={
        enhanced,
        parbox=false,boxrule=0mm,boxsep=0mm,arc=0mm,
        outer arc=0mm,left=5mm,right=3mm,top=7pt,bottom=7pt,
        toptitle=1mm,bottomtitle=1mm}}

\newtcolorbox{noteBox}{textmarker,
    borderline west={4pt}{0pt}{gray},
    colback=gray!10!white}
\newcommand{\note}[1]{\begin{noteBox} #1 \end{noteBox}}

\newcommand{\orcidlink}[1]{\textsuperscript{\href{https://orcid.org/#1}{\includegraphics[scale=0.2]{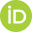}}}}

\begin{document}

\title{How Soft Skills Shape First-Year Success \\in Higher Education} 
\makeatletter
\newcommand{\linebreakand}{%
  \end{@IEEEauthorhalign}
  \hfill\mbox{}\par
  \mbox{}\hfill\begin{@IEEEauthorhalign}
}
\makeatother
\author{\IEEEauthorblockN{Kerstin Andree\orcidlink{0009-0007-6360-8661}}
\IEEEauthorblockA{\textit{Technical University of Munich}\\
Heilbronn, Germany}
\and
\IEEEauthorblockN{Santiago Berrezueta-Guzman\orcidlink{0000-0001-5559-2056}}
\IEEEauthorblockA{\textit{Technical University of Munich}\\
Heilbronn, Germany}
\and
\IEEEauthorblockN{Stephan Krusche\orcidlink{0000-0002-4552-644X}}
\IEEEauthorblockA{\textit{Technical University of Munich}\\
Munich, Germany}
\linebreakand
\IEEEauthorblockN{Luise Pufahl\orcidlink{0000-0002-5182-2587}}
\IEEEauthorblockA{\textit{Technical University of Munich}\\
Heilbronn, Germany}
\and
\IEEEauthorblockN{Stefan Wagner\orcidlink{0000-0002-5256-8429}}
\IEEEauthorblockA{\textit{Technical University of Munich}\\
Heilbronn, Germany}
}

\makeatletter
\renewcommand\@makefnmark{}
\makeatother

\maketitle

\footnotetext{
This is the authors' \textbf{preprint version} of a paper submitted to the 4th IEEE German Education Conference (GECon 2025). The final published version will be available via IEEE Xplore Library.
}

\begin{abstract}

Soft skills are critical for academic and professional success, but are often neglected in early-stage technical curricula. This paper presents a semi-isolated teaching intervention aimed at fostering study ability and key soft skills—communication, collaboration, and project management—among first-year computer science students. The elective seminar \textit{Soft Skills and Tools for Studies and Career in IT} was alongside a mandatory team-based programming course. We analyze project outcomes and student experiences across three cohorts across three groups: students who attended the seminar, students who teamed up with a seminar attendee, and students with no exposure to the seminar. 

Results show that seminar participants performed significantly better in individual presentations and team projects. Qualitative feedback further indicates improved team dynamics and study preparedness. Although self-assessed collaboration and communication did not reach statistical significance, consistent trends suggest that early soft skills training enhances academic integration. We recommend embedding such interventions early in technical study programs to support the transition into university life. 

\end{abstract}


\section{Introduction}\label{sec:intro}

Soft skills encompass social, emotional, and subjective competencies~\cite{Freitas2022}. These skills are applicable across various professional fields and are widely acknowledged as fundamental to academic achievement and career success~\cite{Caeiro2021,tang2019beyond,Karimova2020}. Literature has shown that soft skills should be best developed in higher education~\cite{Chamorro‐Premuzic01032010}. In particular, soft skills training with project-based learning should be integrated to effectively cultivate these abilities among students~\cite{Freitas2022}. 

Key soft skills include presentation, communication,  emotional intelligence, and conflict management~\cite{Otermans2023}. Students should record and reflect on presentations to enhance self-awareness and strengthen leadership and emotional intelligence. Role-playing exercises have also proven effective in fostering emotional intelligence, empathy, and leadership capabilities.
Additionally, listening skills should be practiced. By enhancing communication abilities, students can also refine their conflict management skills and finally increase their \textit{study ability}. Study ability is essential for successfully completing a university degree~\cite{Boettcher2024}. It encompasses soft skills, such as effective learning strategies, appropriate communication within the academic environment, and a solid understanding of university structures and codes of conduct. These elements help students navigate their studies with greater confidence.

Teaching soft skills at the beginning of a study program also helps avoid the typical pattern of dropping out due to the first-semester shock~\cite{Boettcher2024}. 
However, despite their importance, studies indicate that graduates often perceive a gap in teaching soft skills within higher education institutions, underscoring the need for more structured training in this area~\cite{tang2019beyond, Otermans2023}.

This paper presents a teaching concept designed to address the need for soft skills training, particularly in higher education in computer science. We developed the seminar \textit{Soft Skills and Tools for Studies and Career in IT}, which was first offered as a support elective in the winter semester, 2024, targeting first-year bachelor's students. The seminar, called here the soft skills seminar, consists of eight learning units covering various aspects of soft skills through interactive and practical exercises, such as presentation skills, project management, and stress resilience. Additionally, the seminar discusses the responsible usage of AI in studies and introduces helpful tools. 

This paper examines whether students truly benefit from the soft skills teaching concept and whether these benefits can be observed from the beginning of their studies. To do so, we analyze the extent to which the performance and quality of project outcomes differ between students who attended the soft skills seminar and those who did not, within the parallel course \textit{Fundamentals of Programming} (FoP). 
FoP is a first-semester bachelor's course designed to give students hands-on experience with programming. As part of the course, students develop a game in small teams and are responsible for managing the study project independently. A previous work~\cite{berrezueta2024code} has revealed that students face several challenges, particularly regarding communication and collaboration skills. Thus, this project-based course provides an ideal setting to evaluate the impact of the taught soft skills.

To investigate the impact of the soft skills seminar on student outcomes in the FoP project and presentation, we formulate the following research questions:

\begin{enumerate}[label=\textbf{RQ\arabic*}, leftmargin=*] 
    \item How does participation in the soft skills seminar influence students' individual performance during the final presentation in the FoP project?
    \item What impact does participation in the soft skills seminar have on teams' performance in the development phase of the FoP project?
\end{enumerate}

To address these research questions, we conduct a statistical analysis comparing the performance in FoP of (1) students who attended both the soft skills seminar and FoP, (2) students who did not attend the seminar themselves but were part of a FoP project team that included at least one student who did, and (3) those who neither attended the seminar nor worked in a team with a seminar participant.

Our results show that students who attended the soft skills seminar significantly performed better in the first-semester study project than students who did not. The data indicates that the key aspects of our soft skills seminar were already implemented at the beginning of the study program. Furthermore, qualitative analyses reveal that the seminar content helped considerably in managing the programming project. 

The paper is structured as follows: Section~\ref{sec:relatedWork} discusses related work. Background knowledge about the two courses is introduced in Section~\ref{sec:background}. Section~\ref{sec:method} explains the methodology of the conducted analysis. Results are presented in Section~\ref{sec:results} and discussed in Section~\ref{sec:discussion}. Section~\ref{sec:conclusion} concludes the paper.

\section{Related work}\label{sec:relatedWork}

There are different approaches to teaching soft skills in higher education. 
Since teaching soft skills is often overlooked in software engineering programs, Pulko and Parikh~\cite{pulko2003teaching} propose a semi-isolated approach. \textit{Study skills} are taught in an extra-curriculum course of four two-hour sessions. The course covers presentation skills, teamwork, project management, and report writing. However, key components like self-management, learning techniques, and university codes of conduct are missing. While students consider the course valuable, an analysis of the benefits throughout the study program is missing.

Lamp et al.~\cite{LampKU96} present a teaching approach that integrates soft skills into courses with practical components in each semester of the degree program. As part of this concept, students attend weekly session on soft skills in the form of guest lectures or workshops, with the aim of applying what they learn directly in their hands-on courses. This revised curriculum was tested on a single cohort. However, challenges emerged, including the increased workload for academic staff—many of whom lacked training in this area—as well as maintaining a balance between soft skill development and subject-specific content. While soft skills are seen as essential for both career and academic success, they should not overshadow the development of technical expertise. 

Warrner~\cite{warrner2021integrating} suggests integrating soft skills only into two courses and incorporating communication skills spcifically required for academic environment, e.g., e-mail etiquette. 

Team projects are a common approach to integrating soft skills into study programs~\cite{YOUNIS2021151,gonzalez2011teaching}. While students work one semester on a project, they acquire soft skills by experiencing communication with customers, conflicts within the team, and presenting their final product. Instructors guide and support the teams in each situation. Since these projects are usually scheduled for the last semesters of a study program, this teaching concept does not target the problem of missing study ability in early semesters, leading to an increased probability of drop-outs. Moreover, theoretical input, especially on conflict management and resilience, is missing to better prepare students for internships~\cite{Almeida2023}.

With the proposed concept in this paper, we can consider the soft skills seminar as a semi-isolated course focusing on improving the study ability of first-semester students. Unlike teaching soft skills solely through project-based courses, we offer a solid theoretical foundation supported by lecture slides designed to serve as a practical handbook throughout the study program and in future projects. By introducing methods and techniques early on, we aim to build students' awareness before they face challenging situations. We believe that when students recognize potential difficulties in advance and are equipped with strategies to address them, they are more likely to handle those challenges effectively.

\section{Course Setup}\label{sec:background}
\subsection{Soft Skills Seminar}
The soft skills seminar is offered as a support elective, primarily targeting first-semester students transitioning from school to university. Since universities demand higher levels of self-management, organization, and presentation skills, the course is designed to provide a smooth entry into academic life. Its main goal is to enhance students’ study ability, thereby increasing their chances of successfully completing their degree. The learning goals encompass the development of skills in presentations, communication, project management, resilience, and conflict management. The seminar is divided into eight learning units consisting of an input and an exercise session following the principles of interactive teaching~\cite{krusche2017interactive}.

While input sessions introduce key concepts with integrated smaller student tasks, exercise sessions focus on applying these methods and techniques through hands-on team tasks that have been proven to be effective in teaching soft skills, such as role-plays~\cite{jackson2011teaching}, cooperative learning approaches~\cite{loh2020unravelling}, and interactive games (e.g., tangram to practice active listening skills~\cite{lev2020interacting}). Students are also required to do a presentation in pairs and provide structured, constructive feedback on their peers' presentations. 

Table~\ref{tab:units} provides an overview of the individual learning units. The highlighted units are directly related to the programming course held in parallel. These units were scheduled at the early stage of the study project the students must accomplish in the programming course. For each learning unit, slides were shared with the students to provide a handbook that can be used throughout the study program. Following the principle of constructive alignment~\cite{biggs2003aligning}, the form of assessment is a learning portfolio in which students summarize, visualize, and reflect on their learning outcomes.

\begin{table}[h]
    \centering
     \caption{Overview of soft skills learning units}
    \begin{tabular}{p{0.25cm}p{7cm}}
    \hline
        \textbf{Unit} & \textbf{Topic}  \\ \hline
        01 & Code of Conduct, communication channels and etiquette \\ 
        \rowcolor{gray!20} 02 & Presentation Skills \\
        03 & Helpful tools: LaTeX, literature management, note taking \\
        04 & Stress management \& learning techniques \\
        \rowcolor{gray!20} 05 & Project management \\
        \rowcolor{gray!20} 06 & Teams \& conflict management \\
        \rowcolor{gray!20} 07 & AI in Studies \\
        08 & Entrepreneurship, Startups \& Co \\ \hline
    \end{tabular}
    \label{tab:units}
\end{table}

\subsection{Fundamentals of Programming Course}

Fundamentals of Programming (FoP) is an introductory, hands-on course for first-semester students in the Bachelor's program. No prior programming knowledge is required. The curriculum covers essential topics such as control structures, data types, Object-Oriented Programming (OOP) concepts, streams, graphical user interfaces (GUIs), and recursion. 
Its assessment is fully based on practical activities. As shown in Table~\ref{tab:grades}, 60\% of the final grade comes from developing and presenting a final project.

\begin{table}[h]
    \centering
     \caption{Overview of the grading schema for FoP}
    \begin{tabular}{p{5cm}p{2.5cm}}
    \hline
        \textbf{Assignment} & \textbf{Percentage grade}  \\ \hline
         Homework exercises & 20~\%\\ 
         In-class exercises & 10~\%\\
         In-class weekly quizzes & 10~\%\\
        \rowcolor{gray!20} \textit{Project development} & 40~\%\\
        \rowcolor{gray!20} \textit{Project presentation} & 20~\%\\
        \hline
        \textbf{Total grade} & \textbf{100}~\% \\ \hline
    \end{tabular}
    \label{tab:grades}
\end{table}

The project centers on creating a classic-style 2D game inspired by the concept of \textit{Maze Runner}. The gameplay revolves around steering a character through a complex maze. The main goal is to progress from the starting point to the exit while dealing with hidden traps, hostile entities, and a locked gate that can only be opened with a key. The maze features walls, forming an intricate network of paths, dead-ends, and twists. Players must carefully navigate the environment, evade or confront threats placed throughout the map, and find keys necessary to unlock the exit and complete the level.

Students have two and a half months to complete the project, working in teams of three members. Each team can choose a tutor (a student assistant for the course). While we provide a problem description, a Java project template, a continuous integration (CI) tool, and guidance on using Git for version control, no formal training is given in areas such as task coordination, conflict resolution, collaborative project management, or use of artificial intelligence (AI) tools. The achievement of this project (game working) will give the team a grade between 0 and 100 points, representing 40\% of the final grade in the lecture.  

To qualify for the project presentation, the game must be fully functional, executable, and playable. Instructors first evaluate the project to determine whether it meets the minimum requirements for presentation. The presentation consists of a preliminary component in which students must submit a recorded screencast showcasing their game and explaining how it was developed and solved. We provided detailed guidelines regarding the expected content and set a maximum duration of 8 minutes. Participation in the screencast is flexible—students may choose whether or not to appear on camera and whether all team members are involved in the video.

\section{Methodology}\label{sec:method}

\subsection{Sample Selection}

The total sample comprises 138 first-semester students who participated for the first time in the lecture FoP, forming 46 groups of three members. 

\subsubsection{Groups for Individual Analysis}

The full sample was divided into three groups based on exposure to the soft skills seminar. The \textbf{experimental group} comprised 22 students who personally attended the seminar. \textbf{Control group 1} included 25 students who did not participate in the seminar themselves but were part of a team in which at least one member had attended. Finally, \textbf{control group 2} comprised 91 students who belonged to teams where none of the members participated in the seminar.

The imbalance in the groups, especially control group 2, is because, unlike the FoP course, the soft skills seminar is a support elective and not part of the mandatory curriculum. Additionally, these courses are taught by different instructors and follow separate lecture formats.

\subsubsection{Groups for Team-based Analysis}

We also divided the sample into two team-based groups. The \textbf{experimental team group} included teams in which at least one member participated in the seminar, while the \textbf{control team group} comprised teams where no members attended the seminar. There is an imbalance in the size of the groups; however, the teams were formed based on the student’s preferences, which reflects a natural grouping process rather than an imposed structure. 

\subsection{Data Acquisition}

The data collected in this study consists of the individual grades awarded for the final project presentation, the team grades obtained during the project development phase, qualitative data taken as quotes directly from the learning portfolios, and the results of the mandatory course evaluation and the final FoP survey.

The mandatory course evaluation was conducted during the first half of January, signaling the end of the lecture period. Students enrolled in the soft skills seminar participated in it. The qualitative data was mainly used to assess how students perceived the course’s relevance and the effectiveness of its overall concept.

The FoP survey was conducted at the end of the semester—immediately following the project presentations. To ensure independent and unbiased responses, students were not allowed to discuss their answers during the process. This survey aimed to gather qualitative insights into their experiences throughout the project development phase.
Students rated key aspects of their teamwork experience, including communication among team members, task distribution, and communication with their selected tutor, using a 1–10 scale, where 1 indicated "very poor" and 10 indicated "excellent."

\subsection{Statistical Analysis}

For the individual analysis, we computed descriptive statistics for each group, including the sample size (N), the average presentation grade (Mean), the grade dispersion (Standard Deviation), as well as the minimum and maximum values. These results are summarized in Table~\ref{Statistics} and visually represented in Figure~\ref{Boxplot} for better interpretation.
Given the presence of two control groups and one experimental group, we conducted a Tukey Honestly Significant Difference (HSD) test to identify which specific group comparisons showed statistically significant differences. The results of this post-hoc analysis are presented in Table~\ref{HSD}.

For the team-based analysis, we applied the same statistical procedures—descriptive statistics and HSD test—as in the individual analysis to ensure consistency and comparability of results. These results are summarized in Table~\ref{TeamStatistics} and are visually represented in Figure~\ref{BoxplotTeam-based} for more precise interpretation.

The course evaluation analysis does not provide statistical results but provides qualitative results. Finally, the FoP survey presents meaningful statistical results summarized in Table \ref{tab:survey_stats_test} and illustrated in Figure \ref{Boxplot-Survey}.

\section{Results}\label{sec:results}

\subsection{Individual Analysis Results}

After processing the individual presentation grades, Figure~\ref{Boxplot} illustrates that the experimental group achieved a high mean score with minimal variability, suggesting a consistent and positive effect of the soft skills seminar. Control group 1 also demonstrated a relatively high median score but with greater dispersion compared to the experimental group. In contrast, control group 2 exhibited substantial variability (cf. Table~\ref{Statistics}: standard deviation = 15.95), including several low outliers—some as low as 0\%—which may reflect the absence of seminar-related benefits within this group.

\begin{figure}[h]
    \centering
    \includegraphics[width=1\linewidth]{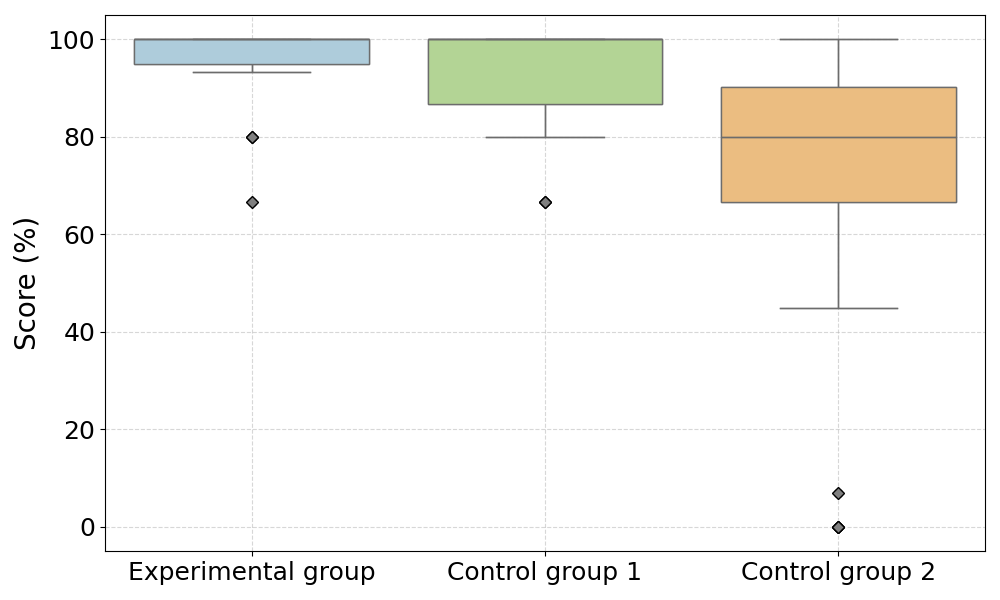}
    \caption{Comparison of performance scores across the experimental group, control group 1, and control group 2.}
    \label{Boxplot}
\end{figure}

\begin{table}[ht]
\centering
\caption{Descriptive statistics comparison }
\begin{tabular}{l c >{\columncolor{gray!20}}c c c c}
\hline
\textbf{Group} & \textbf{N} & \textbf{Mean} & \textbf{Std. Dev.} & \textbf{Min}  & \textbf{Max} \\
\hline
Experimental group  & 22 & \cellcolor{green!15}95.21 & 8.46  & 66.70 & 100.00 \\
Control group 1      & 25 & 90.27 & 9.83  & 66.70 & 100.00 \\
Control group 2      & 91 & 82.20 & \cellcolor{red!20}15.95 & \cellcolor{red!20}0.00 & 100.00 \\
\hline
\label{Statistics}
\end{tabular}
\end{table}

Table~\ref{HSD} indicates that the comparison between the experimental group and control group 2 revealed a statistically significant difference (p \textless{} 0.05), supporting the effectiveness of the seminar. In contrast, the comparison between the experimental group and control group 1 did not yield a statistically significant difference, suggesting that being part of a team with at least one member who attended the seminar may have conferred some indirect benefits.

\begin{table}[ht]
\centering
\caption{Tukey HSD post-hoc comparison between groups}
\begin{tabular}{l l c c c}
\hline
\textbf{Group 1} & \textbf{Group 2} & \textbf{Mean Diff.} & \textbf{p-adj} & \textbf{Significant} \\
\hline
Exp. group & Control group 1 & 3.486 & 0.7656 & \cellcolor{green!15}False \\
Exp. group & Control group 2 & 14.359 & 0.0016 & \cellcolor{red!15}True \\
Control group 1 & Control group 2 & -10.872 & 0.0154 & True \\
\hline
\label{HSD}
\end{tabular}
\end{table}

\subsection{Team-based Analysis Results}

After processing the team grades, Figure~\ref{BoxplotTeam-based} presents a comparison between the experimental and control team-based groups in terms of final project scores. The experimental group, composed of teams with at least one member who attended the seminar, demonstrates a higher median score and less variability in performance. This indicates a consistent and generally stronger performance across these teams.

In contrast, the control group, made up of teams with no seminar participants, displays greater variability, with a wider interquartile range and a lower median score. Notably, this group also includes several low outliers, with team scores dropping to as low as 50. These findings suggest that exposure to the seminar—either directly or indirectly—influenced team outcomes positively, leading to both improved performance and more uniform results within the experimental group.

\begin{figure}[h]
    \centering
    \includegraphics[width=1\linewidth]{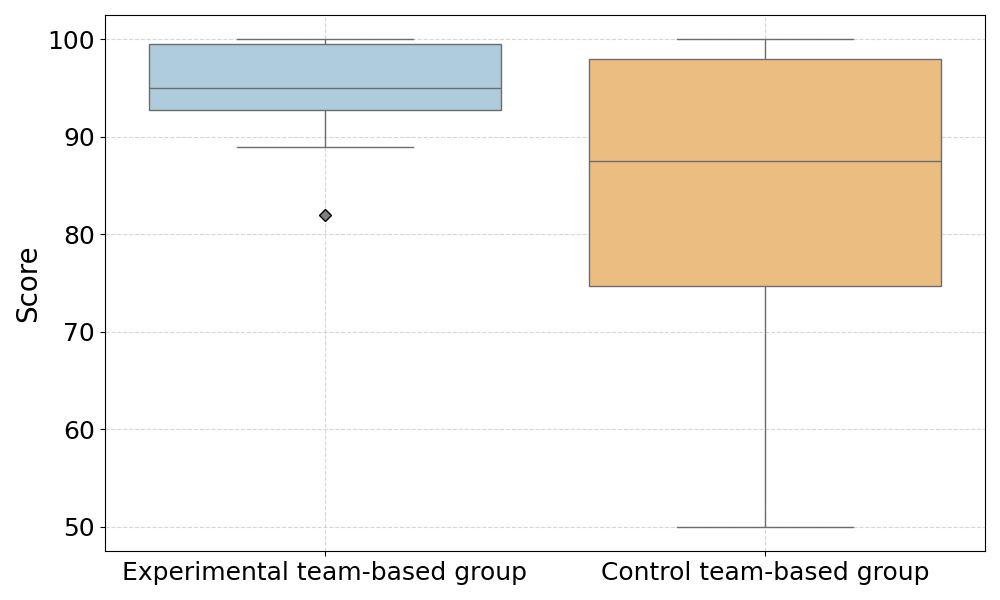}
    \caption{Comparison of performance scores across the experimental team group and control team group.}
    \label{BoxplotTeam-based}
\end{figure}

Table~\ref{TeamStatistics} summarizes the descriptive statistics for the experimental and control team-based groups. The experimental team group (N = 15), which included at least one seminar participant per team, achieved a higher average score (M = 95.13) with relatively low variability ( SD = 5.02). In contrast, the control team group (N = 32), composed of teams without any seminar participants, had a lower average score (M = 85.42) and greater score dispersion (SD = 13.38), with scores ranging from 50.0 to 100.0.

\begin{table}[ht]
\centering
\caption{Descriptive statistics for the team-based groups.}
\begin{tabular}{l c >{\columncolor{gray!20}}c c c c}
\hline
\textbf{Group} & \textbf{N} & \textbf{Mean} & \textbf{Std. Dev.} & \textbf{Min}  & \textbf{Max} \\
\hline
Exp. team group  & 15 & \cellcolor{green!15}95.13 & 5.02  & 82.0 & 100.00 \\
Control team group 1      & 32 & 85.42 & 13.38  & 50.0 & 100.00 \\
\hline
\label{TeamStatistics}
\end{tabular}
\end{table}

To evaluate whether this observed difference was statistically significant, we performed a Tukey HSD post-hoc test. The results revealed a mean difference of 9.71 points between the two groups, with a p-value of 0.0095. This result is statistically significant (p \textless{} 0.05), indicating that teams exposed to the seminar—through at least one member—performed significantly better than those without any exposure.

\subsection{Qualitative Analysis}

Qualitative data was extracted from the learning portfolio submitted by the students who attended the soft skills seminar, the general course evaluation of the seminar, and the survey conducted after the presentations of the FoP project. 

Students emphasized the seminar's usefulness in the FoP project context. One student mentioned: “The techniques shown in the presentation have helped my team to do very well in our FoP project. By following the core lessons in this learning unit, we completed our group tasks efficiently and avoided any internal conflicts, which was truly a miracle.” Another shared how specific lessons on team conflict management were instrumental: “[Being rated as one of the best teams] would not have been possible without learning the core concepts of proper team management from the [soft skills] seminar and how to preserve harmony and productivity in the team, so I thank you from the bottom of my heart for that!”.

Additional feedback revealed that students applied learned methods and integrated tools introduced during the seminar to better organize their projects. Even those who dropped out of the seminar for personal reasons reported that the materials provided supported them in their FoP projects. The concept of the seminar itself also received strong endorsement, with a student commenting: “I believe this class is the perfect support elective module for any first-semester students [...] as it guides them straight away on how to survive university life.” 

In general, students attending the soft skills seminar highly recommend the course and think it is relevant to their studies and future careers. Especially the structure and content of the seminar were positively highlighted, indicating that the course concept also works from a didactic point of view.

\subsection{FoP Survey Analysis Results}

After analyzing the results of the FoP final survey, Figure~\ref{Boxplot-Survey} shows that the experimental group reported consistently higher median scores and less variability across all three survey dimensions: communication among team members, task distribution, and communication with the tutor. In contrast, the control group responses were more dispersed, with several low outliers—particularly in team communication and tutor interaction—indicating a more uneven experience. These visual trends suggest a positive influence of the soft skills seminar on team dynamics and communication. 

\begin{figure}[h]
    \centering
    \includegraphics[width=1\linewidth]{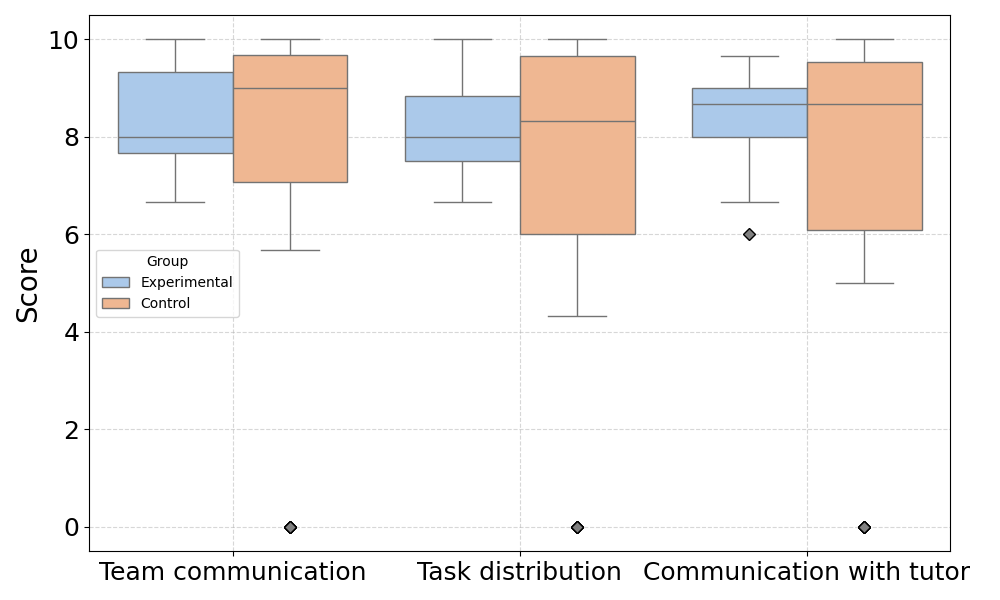}
    \caption{Comparison of performance scores across the experimental team group and control team group.}
    \label{Boxplot-Survey}
\end{figure}

However, as shown in Table~\ref{tab:survey_stats_test}, statistical tests revealed that the differences between groups were not significant. While the experimental group showed higher average scores in all categories, the differences did not reach statistical significance, with the largest gap observed in communication with the tutor (M\textsubscript{exp} = 8.33 vs. M\textsubscript{ctrl} = 7.26; p = 0.14). This suggests a promising trend that aligns with qualitative observations, but further studies with larger samples are needed to confirm the effects quantitatively.

\begin{table}[ht]
\centering
\caption{Statistical comparison between experimental team group and control team group.}
\begin{tabular}{l c c c c}
\hline
\textbf{Measure} & \textbf{\makecell{Exp \\mean/Std}} & \textbf{\makecell{Ctrl\\ mean/Std}} & \textbf{p-value} & \textbf{Significant} \\
\hline
\makecell{Team\\communication} & 8.40 / 1.16 & 7.52 / 3.33 & 0.23 & \cellcolor{red!15}False \\
\makecell{Task\\distribution} & 8.24 / 1.02 & 7.19 / 3.34 & 0.14 & \cellcolor{red!15}False \\
\makecell{Communication\\with tutor} & 8.33 / 1.15 & 7.26 / 1.49 & 0.14 & \cellcolor{red!15}False \\
\hline
\end{tabular}
\label{tab:survey_stats_test}
\end{table}

%TODO: include results of teaching evaluation: (I included it a bit in qualitative without mentioning the numbers :D) I hope that's fine
% structure and content of the seminar (1 best, 5 worst): (1.5+1.2+1.5) / 3 =
% 100% of the participants (n=12) recommend the seminar and think the seminar is relevant
% overall grade for seminar: 1.2 (n=12)
% participation: 1.8 (1 = always, 5 = never)

\section{Discussion}\label{sec:discussion}

Study ability is a critical factor for academic success. It should be cultivated from the beginning of a student's academic journey to prevent the first-semester shock in students entering higher education. 
Based on the results, we can identify the following findings.

\note{\textbf{Finding 1: Improved Presentation Performance.} Students who attended the soft skills seminar (experimental group) achieved higher average presentation grades (M = 95.21) than both control groups. }

The difference between the experimental group and control group 2 (students in teams with no seminar participants) was statistically significant (p = 0.0016), confirming the positive impact of the seminar on individual presentation performance. We explain this observation by integrating the learning unit on presentation skills into the soft skills seminar including the mandatory exercise of holding a presentation in front of the classroom, and the peer-feedback. Students attending the soft skills seminar were aware of the expectations of a good presentation; additionally, they received feedback before they started the preparation phase for the project presentation in the context of FoP. Thus, they could further enhance their presentation skills. 

\note{\textbf{Finding 2: Indirect Team Benefit.} Students who did not attend the seminar but worked with someone who did (control group 1) performed similarly to the experimental group, with no significant difference (p = 0.7656). }

Finding 2 suggests that team collaboration with seminar participants may confer indirect benefits, such as better project organization or presentation preparation. This is also supported by the fact that control group 2 showed the lowest mean score (M = 82.20), the highest variability (SD = 15.95), and even outliers as low as 0\%, indicating that no exposure to the seminar (direct or indirect) correlates with lower and less consistent performance.
We reason this observation by including the learning unit on team management. By introducing the Belbin Team Roles~\cite{belbin2010management}, the seminar helped students understand that a key factor for team success is having a diverse group — including at least one member responsible for effective team and project management. %Due to a lack of sufficient data, it is not possible to determine with certainty whether the student who also attended the soft skills seminar assumed this role and thus made a substantial contribution to the quality of the project work.
%This reasoning might also explain the third finding.

\note{\textbf{Finding 3: Enhanced Team-Based Performance.} At the team level, groups with at least one seminar participant (experimental team group) scored significantly higher than teams without any participants.}

The Tukey HSD test confirmed this difference as statistically significant (p = 0.0095), showing that seminar exposure positively influenced collaborative work as well.

Overall, these findings indicate that the seminar contributed to an improvement in the study ability of first-semester students, particularly in presentation, communication, project management, and teamwork skills. We explain this improvement with two key aspects covered by the seminar: first, the early opportunity for social bonding and peer support through group activities in exercise sessions, and second, the direct application of techniques and methods, which students reported implementing in their FoP projects. 

The coordinated course schedule made it possible to adequately prepare the students for the project work. This observation is consistent with the research literature, which recognizes the advantage of teaching soft skills directly in the context of a project while acknowledging the benefits of semi-isolated courses.

Nevertheless, this study has some limitations. The findings are based on a small sample size due to a relatively small group of students registered for the soft skills seminar. Moreover, the observations were limited to a single course of the first semester, which restricts the generalizability of the results.

\section{Conclusion}
\label{sec:conclusion}

This paper introduced a new approach to teaching soft skills in higher education through a semi-isolated course format. The course's design centers on providing students with theoretical foundations and materials aiming to equip students with models, methods, and techniques that prepare them for the challenges of academic life. 
To investigate the impact of a semi-isolated soft skills course on study ability at the beginning of a study program, we conducted a statistical analysis comparing the performance of student teams in a concurrent programming course project—specifically between teams that included at least one member who attended the soft skills seminar and teams with no attendees.

The results of our study indicate a positive impact of this seminar on the programming project conducted in the first semester. 
Students feel better prepared for academic life and appreciate having a handbook of guidelines to address challenges throughout their study program, e.g., presentations, communication, and teamwork. Thus, we minimize the probability of students suffering from first-semester shock while increasing the students' chances of successfully graduating from the program.

We conclude that offering a semi-isolated soft skills course at the beginning of the study program provides students with essential skills and support right from the start, helping them navigate the challenges of university life more effectively. Therefore, we propose incorporating a combination of these two approaches—semi-isolated and project-based integration of soft skills—into the curriculum of study programs.

We plan to scale the seminar and evaluate its impact across larger sample sizes for future work. Additionally, extending the study to different courses within the first semester will enhance the analysis of increased study ability. Conducting a study throughout the entire program, including higher semesters, would allow for assessing soft skills development in more complex contexts such as advanced presentation settings or scientific report writing.

\balance
\bibliographystyle{ieeetr}
\bibliography{Paper}

\end{document}